\documentclass[preprint,12pt,3p]{elsarticle}
\usepackage{float}
\floatstyle{plaintop}
\restylefloat{table}
\usepackage{graphicx}
\usepackage{booktabs,makecell}
\usepackage{multirow}
\usepackage{pdflscape}
\usepackage{caption}
\usepackage{hyperref}
\usepackage{rotating}
\usepackage{subfigure}
\usepackage[export]{adjustbox}
\usepackage{comment}
\usepackage{amsmath}
\usepackage{longtable}
\usepackage{algorithm}
\usepackage[noend]{algpseudocode}
\usepackage{siunitx}
\usepackage{tikz}
\usepackage{xcolor}
\usetikzlibrary{positioning}
\usepackage{placeins}
\usepackage{graphics}
\usepackage{xcolor}
\usepackage{epstopdf}
\AppendGraphicsExtensions{.tif}
\usepackage{amssymb}
\newcolumntype{M}{>{\centering\arraybackslash}m{3cm}}

\usepackage{enumitem}
\usepackage{array}
\newcolumntype{P}[1]{>{\centering\arraybackslash}p{#1}}

\newcommand{\PreserveBackslash}[1]{\let\temp=\\#1\let\\=\temp}
\newcolumntype{C}[1]{>{\PreserveBackslash\centering}p{#1}}
\newcolumntype{R}[1]{>{\PreserveBackslash\raggedleft}p{#1}}
\newcolumntype{L}[1]{>{\PreserveBackslash\raggedright}p{#1}}

\begin{document}
\begin{frontmatter}

\title{Deep Learning for Brain Age Estimation: A Systematic Review}
\author[label1]{M. Tanveer}
\ead{mtanveer@iiti.ac.in}
\author[label11]{M. A. Ganaie}
\ead{mudasirg@umich.edu}
\author[label2]{Iman Beheshti}
\ead{Iman.beheshti@umanitoba.ca}
\author[label4]{Tripti Goel}
\ead{triptigoel@ece.nits.ac.in}
\author[label1,label20]{Nehal Ahmad}
\ead{nehalahmad@iiti.ac.in}
\author[label5]{Kuan-Ting Lai}
\ead{ktlai@ntut.edu.tw}
\author[label6]{Kaizhu Huang}
\ead{kaizhu.huang@dukekunshan.edu.cn}
\author[label7]{Yu-Dong Zhang}
\ead{yudongzhang@ieee.org}
\author[label8,label8b]{Javier Del Ser}
\ead{javier.delser@tecnalia.com}
\author[label9]{Chin-Teng Lin}
\ead{Chin-Teng.Lin@uts.edu.au}
\address[label1]{Department of Mathematics, Indian Institute of Technology Indore, Simrol, Indore, 453552, India}
\address[label11]{Department of Robotics, University of Michigan, Ann Arbor, MI 48109, USA}
\address[label2]{Department of Human Anatomy and Cell Science, Rady Faculty of Health Sciences, Max Rady College of Medicine,
University of Manitoba, Winnipeg, MB, Canada}
\address[label4]{Biomedical Imaging Lab, National Institute of Technology Silchar, Assam 788010, India}
\address[label20]{ Department of Electrical Engineering and Computer Science, National Taipei University of Technology, Taipei, Taiwan}
\address[label5]{Department of Electronic Engineering, National Taipei University of Technology, Taipei, Taiwan}
\address[label6]{Data Science Research Center, Duke Kunshan University, China}
\address[label7]{School of Computing and Mathematical Sciences, University of Leicester, UK}
\address[label8]{TECNALIA, Basque Research and Technology Alliance (BRTA), 48160, Derio, Spain}
\address[label8b]{University of the Basque Country (UPV/EHU), 48013, Bilbao, Spain}
\address[label9]{School of Computer Science, Faculty of Engineering and Information Technology, University of Technology Sydney, Sydney, Australia}

\begin{abstract}
Over the years, Machine Learning models have been successfully employed on neuroimaging data for accurately predicting brain age.   Deviations from the healthy brain aging pattern are associated to the accelerated brain aging and brain abnormalities. Hence, efficient and accurate diagnosis techniques are required for eliciting accurate brain age estimations. Several contributions have been reported in the past for this purpose, resorting to different data-driven modeling methods. Recently, deep neural networks (also referred to as \emph{deep learning}) have become prevalent in manifold neuroimaging studies, including brain age estimation. In this review we offer a comprehensive analysis of the literature related to the adoption of deep learning for brain age estimation with neuroimaging data. We detail and analyze different deep learning architectures used for this application, pausing at research works published to date quantitatively exploring their application. We also examine different brain age estimation frameworks, comparatively exposing their advantages and weaknesses. Finally, the review concludes with an outlook towards future directions that should be followed by prospective studies. The ultimate goal of this paper is to establish a common and informed reference for newcomers and experienced researchers willing to approach brain age estimation by using deep learning models.
\end{abstract}


\end{frontmatter}

\section{Introduction}

With the global rise of population, a spike of cases has been noted worldwide in a range of non-fatal albeit disabling disorders, including neurodegenerative disorders such as cognitive decline and dementia \cite{vos2012years}. To deal with this challenge, there is a growing need to identify the link between brain aging processes and neurodegenerative disease mechanisms. Methods need to be devised for identifying the subjects who are at a higher risk of age-associated deterioration, monitoring the progress of degradation, and devising appropriate treatments to be prescribed for involved subjects. Age-related brain alterations play an important role in the etiology of brain diseases. The discrepancy between the brain age and the chronological age may reveal several risks of experiencing health related issues during the different stages of life.

In this context, an unquestioned upsurge of data-driven models has been developed for human health over the last couple of decades. Indeed, machine learning has been successful in different modeling tasks related to brain age, mostly using magnetic resonance imaging (MRI) scans of the brain. All such tasks fall below the overarching \emph{brain age estimation} concept which, through brain imaging data coupled with machine learning models, encompasses techniques capable of predicting homogeneous trajectories in normal brains at an individual level \cite{franke2019ten}. A brain age estimation framework typically employs a training set of cognitively healthy participants together with supervised learning (i.e., a regression algorithm) to model the correlation between extracted brain features (i.e., independent variables) and the real age of the patient (i.e., dependent variable). This prediction model is then applied to independent test data to infer the brain age of patients unseen by the trained model. The deviation (i.e., positively or negatively) from these typical trajectories, often called the brain age delta (estimated brain age minus real age), have been linked to the status of the brain \cite{franke2019ten}. A large number of studies have documented that a positive brain age delta is associated to brain abnormalities and mortality \cite{cearns2019recommendations}, whereas a negative brain age delta is linked to a younger brain \cite{luders2016estimating}. The brain age metric has been successfully used to detect different neurological and mental diseases \cite{mishra2021review}. Beyond the detection of neurological and mental illnesses, the brain age metric has also been utilized for discovering the effects of lifestyles (such as resilience, depressive symptoms, life satisfaction) on cognitively healthy brains \cite{sone2022neuroimaging}. 
 
It is worthwhile noting that the accuracy of interpreting brain age results relies on the robustness of the brain age estimation framework in use. Indeed, a more accurate brain age estimation framework can deliver more robust outcomes for clinical applicants. As a result, developing a more precise brain age estimation framework for clinical applications is required, so many research groups have attempted at improving brain age estimation frameworks by embracing diverse machine learning techniques. In addition to the traditional machine learning methods, deep learning has more recently become a popular methodology in the field of neuroimaging, where it has been widely employed for different tasks such as segmentation, lesion detection, and classification \cite{zhang2020survey}. One of the most significant advantages of deep learning models is their ability to combine feature extraction, feature reduction, and prediction stages into uniform computational system capable of outperforming traditional machine learning methods, especially when modeling highly complex data. Therefore, deep learning has risen to prominence as a preferred technique for brain imaging studies, and deep learning-based neuroimaging studies have increased with a steadily growing propensity over the last decade \cite{zhang2020survey}.
\begin{table*}[h!]
\caption{Published surveys reviewing brain age estimation studies.}
\label{tab:Published surveys reviewed}
\resizebox{\columnwidth}{!}{\begin{tabular}{cccccC{3cm}cL{9cm}}
\toprule
Reference & Year & \makecell{Reviewed\\period} & \makecell{Papers\\reviewed} & \makecell{Dataset\\Coverage} & \makecell{Models} & \makecell{Deep\\Learning?} & Comments  \\
\midrule
\cite{sajedi2019age} & 2019 & 2010-2018 & 40 & Yes & Traditional, Deep Learning & Yes & Focused on T1w-MRI preprocessing methods, feature extraction techniques, and regression algorithms. A brief discussion of deep learning models (i.e., the CNN model) and associated toolboxes (i.e., TensorFlow and PyTorch) \\
\midrule
\cite{baecker2021machine} & 2021 & 2010-2021 & n.a. & No & Traditional & No & Focused on T1w-MRI preprocessing methods, the construction of a standard brain age study, and clinical applications of brain age prediction \\
\midrule
\cite{mishra2021review} & 2021 & 2010-2021 & 84 & No & Traditional & No & Focused on different brain imaging modalities (T1w-MRI, fMRI, DTI, PET, and SPECT),  feature extraction and reduction methods, regression algorithms, bias adjustment methods, and clinical applications of brain age prediction \\
\midrule
\textbf{Ours} & 2022 & 2017-2022 & 35 & Yes & Deep Learning & Yes & Review of deep learning architectures used in the area, prospective lines of research \\
\bottomrule
\multicolumn{8}{l}{}\\
\multicolumn{8}{l}{\makecell[l]{Legend: T1w-MRI: T1-weighted magnetic resonance imaging; fMRI: functional MRI; DTI: diffusion tensor imaging; PET: Positron\\emission tomography; SPECT: single photon emission computed tomography}}
\end{tabular}}
\end{table*}

To date, only a few surveys have summarized existing brain age estimation studies \cite{sajedi2019age, baecker2021machine, mishra2021review}. Of note, these surveys have mostly focused on traditional machine learning algorithms (such as support vector regression, relevance vector regression, and Gaussian process regression), feature extraction and data reduction techniques, as well as the clinical application of the developed brain age techniques. Table \ref{tab:Published surveys reviewed} shows the list of published surveys that have summarized brain age studies so far. In the light of the information provided in the table and as per our examination of the literature contributed in the are, no survey has previously offered a systematic review of the state-of-the-art related to deep learning for brain age estimation. To the best of our knowledge, the present study is the first that addresses this niche by carefully inspecting different deep learning architectures that have been adopted to estimate the brain age from neuroimaging data. Our detailed literature analysis is further complemented with a proposal of future directions in this area, stimulated by our informed conclusions drawn from the examined literature corpus. Our main contributions can be summarized as follows:
\begin{itemize}
    \item We categorise existing brain age estimation models based on the deep learning architectures used for the estimation. Moreover, we further classify such models depending on whether they adopt a slice-based or a voxel-based approach when modeling their input data. 
    \item We examine in detail such estimation models by discussing about their modalities, size of data in use, reported performance of the models, and the datasets used for evaluation purposes.
    \item We examine every deep learning architecture used in the area, describing design patterns in their different hyperparameters, including the number of hidden layers, activation functions, dense layers, loss functions and other choices alike.
    \item We enumerate several challenges that remain unsolved and trace directions that can be followed for the advancement of data-based solutions for predicting the brain age from image data.
\end{itemize}

The rest of this review is structured as follows. Section \ref{sec:preliminary} describes briefly the deep learning architectures referred later in the literature review. Section \ref{sec:taxonomy} establishes the taxonomy around which the literature examination is done, and reviews the publications that use deep learning architectures for estimating the brain age from neuroimaging data. Finally, limitations of existing studies and suggestions for future research paths are given in Section \ref{sec:challenges}. Finally, Section \ref{sec:conclusions} concludes this review with a summary of the main insights drawn from the survey.
 
\section{Deep Learning Architectures} \label{sec:preliminary}

Before proceeding with the literature review, it is necessary to establish concepts and architectures related to deep neural networks that will be later referred to as such extensively in the bibliographic study. This section provides a brief explanation of some popular deep learning models for the sake of completeness and to ease the understanding of the results of our examination.

\begin{figure*}[h!]
    \centering    \includegraphics[width=1\textwidth]{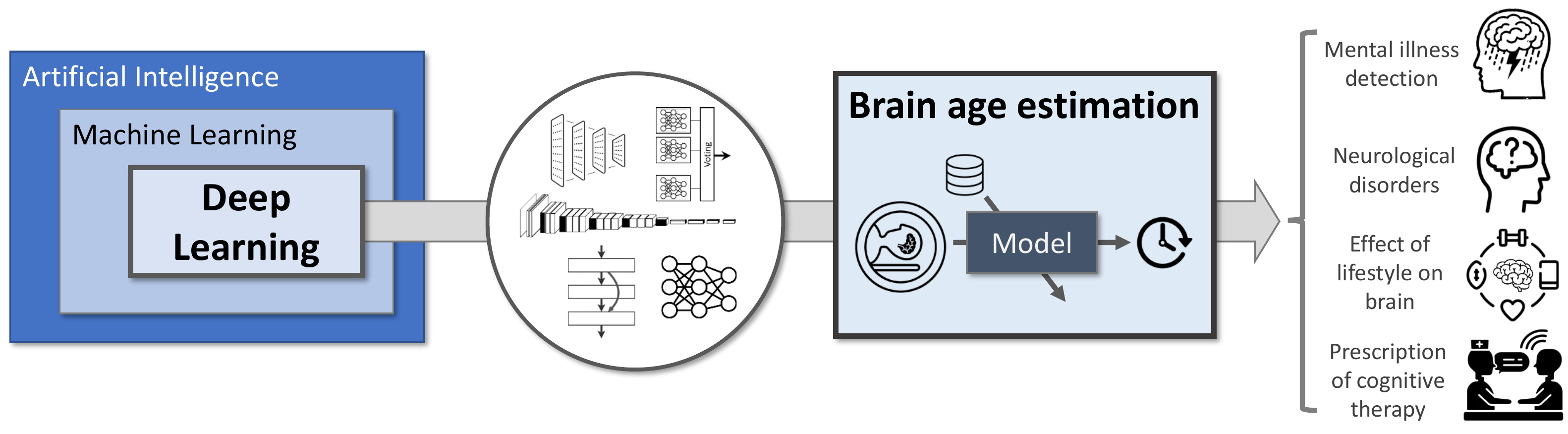}
  \caption{Conceptual diagram relating different Deep Learning architectures with the targeted application (brain age estimation) and the ultimate purposes for which the application is sought.}
  \label{fig:concept}
\end{figure*}

As the technology develops day by day, the amount of global digital information increases exponentially. The medical domain has arguably been among the sectors whose digital substrate has grown sharply over the years. As a result, high-performance models have been in large demand to ingest, process and extract knowledge from these data. For this purpose, the advent and progressive maturity of deep learning models has gained a significant momentum in the majority of the research areas, particularly in biomedical data processing. Deep learning models have comprehensively been proven to advance remarkably in the analysis of biomedical data for different diseases, such as Alzheimer \cite{tanveer2020machine}, Parkinson \cite{tanveer2022parkinson} and brain tumor classification \cite{muhammad2020deep}. Here, we give brief overview of some of these deep learning architectures used in brain age estimation, including  Feed-Forward Neural Networks, Convolutional Neural Network (CNN), VGGNet \cite{simonyan2014very}, ResNet \cite{emmert2020introductory}, and ensembles of Deep Learning models \cite{condorcet1785essay}. Such architectures, the application under focus in this review and the different goals for which brain age can be estimated using Deep Learning architectures are conceptually illustrated in Figure \ref{fig:concept}.

\subsection{Feed-forward neural network}

It is the simplest form of a neural network. The primary objective of a feed-forward neural network is to compute the approximation of a function \cite{goodfellow2016deep}. Feed-forward networks are sequential functions or perceptrons assembled together in a chain structure and it is syndicated with directed acyclic graph representing the way functions are associated together. In this network, data flow in only one direction. The total number of layers in a network represents the \emph{depth} of the network, so that the higher the number of layers in the network is, the \emph{deeper} the network can be declared to be. The parameters defining such composed functions in the network are adjusted to minimize a loss function defined over a training dataset, by using a gradient backpropagation algorithm. 

\subsection{Convolutional Neural Networks, ResNet and VGGNet} \label{ssec:cnn_resnet_vgg}

The progressive advance in feed-forward neural networks eventually yielded CNNs as the next generation of the wide family of neural computation models. CNN is one of the most successful data-based models arising from the computer vision domain \cite{krizhevsky2012imagenet} which, since its inception, has been extrapolated to other complex data domains, such as video or time series modeling. It is widely being used in applications such as natural language processing \cite{zhao2016attention}, voice processing \cite{dahl2011context} or text mining \cite{peng2018large}, to mention a few. In general, the superior performance of CNNs when compared to other artificial neural networks hinges on their capability to automate the feature extraction process from complex multidimensional data, which is realized by extending the aforementioned gradient backpropagation algorithm through layers that model spatial correlation within the data input to the network. Such layers, referred to as \emph{convolutional}, are the first in a stacked hierarchy that, together with pooling layers and fully connected (dense) layers, compose the overall structure of a CNN used for classification or regression tasks.

The fact that vanilla CNN models struggled to achieve good levels of classification accuracy when dealing with comprehensive datasets paved the way for the proposal of new CNN architectures over the years. AlexNet \cite{krizhevsky2012imagenet} was the first CNN-based modern architecture that came into picture, achieving unrivaled performance at the time in the well-known ImageNet classification competition. Theoretically, neural networks with a large depth have a better capability to extract features, potentially improving the model's predictive performance \cite{he2015delving, donahue2014decaf}. However, when dealing with deep convolutional architectures, the chances of having exploding and vanishing gradients become higher, being catastrophic for the model's training process. To address these issues, Microsoft researchers introduced an enhanced CNN architecture coined as ResNet \cite{simonyan2014very, he2016deep}. The ResNet architecture introduced residual blocks and the concept of skip-connection networks. By virtue of skip connections, the activation of a layer is connected to subsequent layers in the hierarchy by skipping a few layers in between, hence forming a residual block. In other words, a ResNet architecture is composed of stacks of residual blocks. The overall accuracy in ResNet is enhanced without loss of accuracy or increase in the training and testing error. 

Another CNN model called VGGNet was developed by \cite{simonyan2014very}, forging thereafter a series of different models characterized by different depths and hyperparametric configurations. VGGNet reduced the memory consumption and model time complexity as it does not use local response normalization algorithm. VGGNet flavors have different channel specifications, supporting from $11$ weight layers (each including $8$ convolution with three fully connected layers) to $19$ weight layers (including $16$ convolution layers with $3$ fully connected layer each). Even after having multiple variants of CNN models, researchers were enthusiastic to enhance the model performance and generalization capability. 

\subsection{Ensemble Deep Learning}

In order to further enhance the performance of the model over unseen query data, ensemble learning soon entered the landscape of deep learning models. In general, ensemble learning involves combining and training multiple models with the aim of improving generalization performance \cite{hansen1990neural, dietterich2000ensemble, ganaie2021ensemble, cao2020ensemble, ju2018relative, gonzalez2020practical}. This performance boost is due to the fact that predictions from different networks modeling diversified knowledge about the training dataset are combined together to give better performance \cite{lee2015m}. Ensemble learning approaches using deep neural networks can be found within the broad categories of bagging \cite{breiman1996bagging}, boosting \cite{bartlett1998boosting} and stacking \cite{wolpert1992stacked}. These networks are considered as the best in terms of performance accuracy and robustness as compared to individual networks \cite{ huang2017snapshot, ren2016ensemble}. Ensemble CNNs are desirable because of the basic fact that the optimization problem of assigning weight to each node may have different local minima \cite{hansen1990neural, sharkey2012combining}. CNN ensembles have so far applied to many modeling tasks, including character recognition, image analysis, face recognition and medical image diagnosis. 

\section{Taxonomy and Literature Review} \label{sec:taxonomy}

Departing from the deep learning architectures revisited above, we now delve into the review of the most significant works for brain age estimation using such architectures. In doing so, the section first settles the methodology followed to collect and organize all the considered corpus of literature (Subsection \ref{ssec:search}), followed by a detailed study of the contributions around a proposed taxonomy that sorts them depending on the deep learning architecture in use (namely, CNN, VGGNet, ResNet, ensemble deep learning, and miscellaneous), and then in terms of the input data given to the model (slice-based or voxel-based. This literature review is given in Subsection \ref{ssec:taxonomy_review}.

\subsection{Search Strategy} \label{ssec:search}

Our literature compilation was performed in March $2022$, querying relevant bibliographic databases (Google Scholar, PubMed) with search keywords tightly connected to the area under study. Specifically, we considered ``deep learning” combined with the following items: ``brain age estimation”, ``brain age prediction”, ``MRI”, ``brain imaging”, and ``neuroimaging”. Studies not focusing on deep learning and neuroimaging were eliminated in the initial screening. A total of $35$ studies were retained and examined in detail. 

\subsection{Taxonomy and Literature Review} \label{ssec:taxonomy_review}

\begin{figure}[h!]
    \centering    \includegraphics[width=0.8\textwidth]{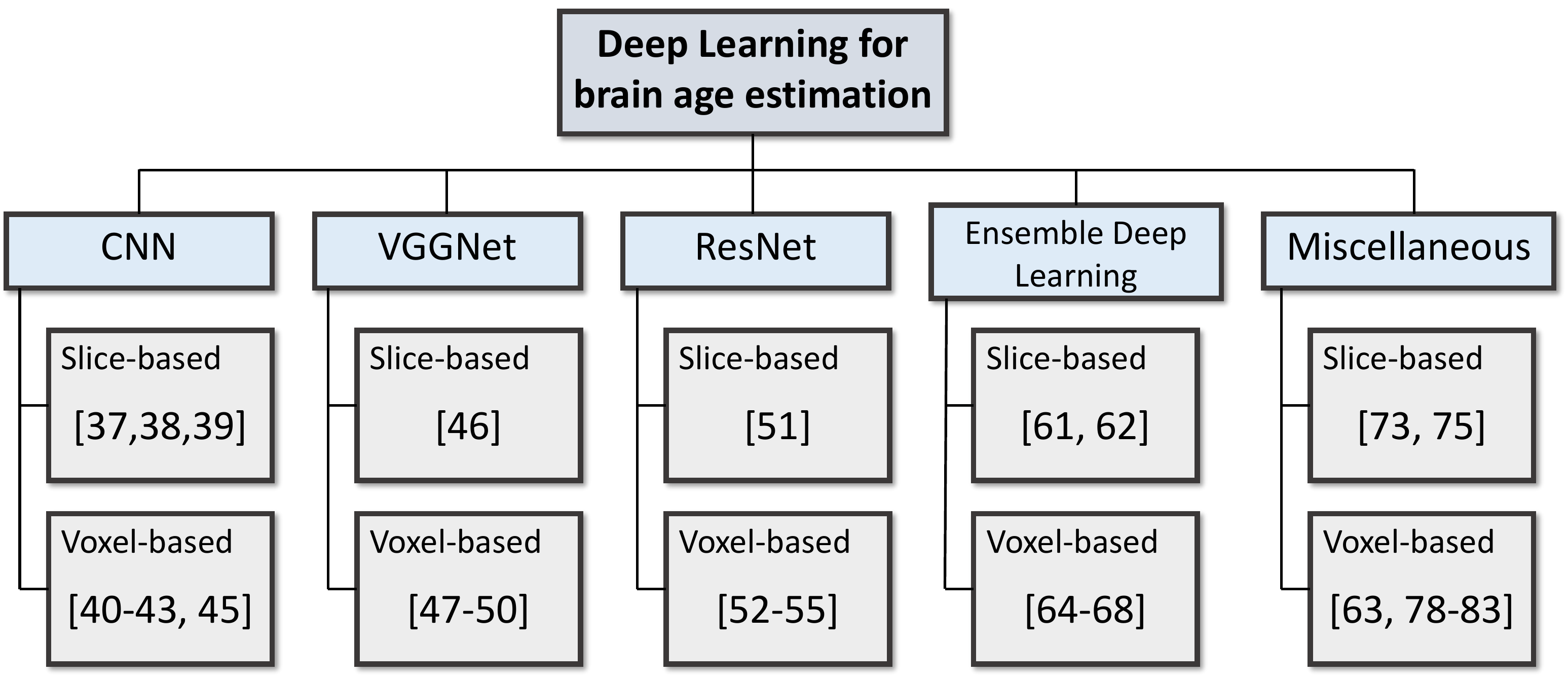}
  \caption{Literature taxonomy of Deep Learning models applied to the estimation of the brain age from neuroimaging data.}
  \label{fig:taxonomy}
\end{figure}

As a result of our literature study, we designed and arranged all works around the taxonomy shown in Figure \ref{fig:taxonomy}. Two classification criteria can be distinguished: the first is the deep learning architecture, which echoes the architectures revised in the previous section and complements it with an additional category (\emph{miscellaneous}) to account for alternative architectural choices not covered in such wide categories. The second classification criterion is the format of the data input to the deep learning model, namely, slice- or voxel-based approaches. In slice-based approaches, deep learning models are trained with the two-dimensional MRI slices extracted from 3D MRI scans. Consequently, slice-based approaches contains less number of trainable parameters, and hence imply a lower computational complexity of the overall modeling solution. However, two-dimensional slices do not contain the whole brain's information (e.g. adjacent correlations between slices are not considered). By contrast, in voxel-based approaches deep learning models are trained using three-dimensional MRI scans or 3D MRI patches. 3D scans are first registered on a standard brain template to make the dimensions of all the 3D MRI scans uniform. After that, the registered 3D scans are fed to a deep learning models for training, wherein the deep learning model is configured with neural processing elements capable of operating over 3D input data tensors. Voxel-based approaches can model brain image data considering inter-slice information. However, the number of trainable parameters is much higher than that of their slice-based counterparts. Bearing this taxonomy in mind, we proceed with our systematic review and analysis of the literature in the subsequent sections, following the two criteria of the taxonomy itself. A summary of the modalities, learning models, datasets in use and reported results of the analyzed studies is given in Table \ref{tab:characteristics}.

\subsubsection{Convolutional Neural Networks}

CNN has attracted researchers since 2017 for brain age estimation and become very popular because of its automatic feature extraction capability and high-performance results. 
\begin{itemize}[leftmargin=*]
\item Slice-based CNN models:
\end{itemize}

When it comes to slice-based CNN approaches, 2D MRI scans are often used to train two-dimensional CNN networks. The work in \cite{lam2020accurate} highlighted the limitations of using 3D CNN for brain age prediction, including the need for a large number of parameters and the computational complexity of the training phase. As a result, it proposed a 2D recurrent neural network (RNN) for brain age prediction. In the proposed model, a 2D CNN encodes important intra-slice features and RNN processes the ordered sequence of sagittal plane MRI scans to learn inter-slice correlations. The weights of the model are initialized using Kaiming weight initialization scheme, whereas training is done by using an Adam solver. The main advantage of the proposed 2D CNN is the reduced number of parameters i.e. $1,068,065$ as compared to $2,018,000$ for 3D CNN with better accuracy. The main limitation of the study is the lack of consideration of the association of brain age with the neurological and other health conditions.

Similarly, a slice-based CNN approach was developed in \cite{amoroso2019deep} by segmenting T$1$-weighted MRI scans to patches. Pearson's correlation is used to measure pairwise similarities between the patches. The extracted metrics are fed to the complex network using the patch as the node and correlation metric as weights. The model shows the most affected brain regions for age prediction using the so-called Gedeon method. They found $10$ most significant features related to aging located in the left hemisphere. The main drawback of the study is the use of a small cohort of only $488$ subjects.

Further along this line, \citet{dular2021improving} recently compared four slice-based CNN models for brain age estimation using transfer learning with domain adaptation and bias correction. The authors generalized the CNN model performance for unseen new MRI data with a simple transfer learning approach. However, the first and fourth CNN models were trained on full resolution $3$D T1-weighted MRI images, whereas the second model was trained on 2D MRI by extracting $15$ axial slices from each 3D MRI image. In the third model, the resolution of the MRI image is downsampled to $95\times79\times78$ for training the 3D CNN model. The input to the different CNN models is of different types, which may affect the performance of the proposed architecture.

\begin{itemize}[leftmargin=*]
\item Voxel-based CNN models:
\end{itemize}

Voxel-based CNN networks are trained over 3D volume MRI brain scans. This is the approach used in \cite{bellantuono2021predicting}, which addressed brain age prediction by introducing complex networks using a structural connectivity model. 3D T$1$-weighted images are segmented into patches and then Pearson correlation is used to calculate the pairwise similarities between the patches. The pair of patches and similarity metric from the graph model is fed to the deep learning model for classification. The developed model demonstrates the effect of aging on different brain regions by exploiting the structural connectivity using a graph model. The proposed model achieved robust results with only brain extraction and linear registration to reduce the computational complexity. Furthermore, to assess the feature importance for age prediction, the Gedeon method is used and represents $12$ most informative patches related to aging. The most informative patches of medial frontal gyrus, caudate, paracentral lobule, putamen, cingulate, brainstem, sub-gyral regions, etc. are observed in Gedeon test. The limitation of the study is the experiments have been performed only on the Autism spectrum disorder dataset, which may not suffice for validating the association of age difference with other disorders.

Another voxel-based strategy was considered in \citet{Cole2017}, where T$1$-weighted images were segmented into gray matter (GM) and white matter (WM) using SPM$12$ toolbox. Results of this study evinced best performance using GM volume compared to WM, raw and ensemble of WM and GM. They also investigated the heritability of the brain age estimation using GM, WM and combination of both. The main limitation of the study is the use of only two scanner data to check the reliability of between-scanner performance of the model. Moreover, the authors did not correlate any neuro-anatomical feature with the predicted age difference, nor did the analysis consider in-scanner motion artefacts that can affect the overall performance of the model in practical settings. Other works along this line include \cite{wang2019gray}, which segmented $1.5$T T$1$-weighted MRI scans into GM, WM and cerebrospinal fluid (CSF). They used GM as the input for training a $3$D CNN network. Logistic regression models and Cox proportional hazard models unveiled the association of age gap with dementia. Attention maps are extracted from the trained networks using gradient-weighted activation mapping, which depicted the changes in GM hyperintensity around the hippocampus and amygdala with the age. On a negative note, the modeling proposal of this work is less generalizable as the proposed CNN network is not able to handle images from different datasets. In addition, the authors excluded dementia and stroke subjects while training the CNN model. 

The hybridization of models has been also explored for the estimation of brain age. An example is the study in \citet{pardakhti2020brain}, proposing to hybridize a 3D CNN with support vector regression \cite{kanwal2018support} and Gaussian process regression \cite{kanwal2018support}. They trained the model using healthy subjects from the IXI dataset, and evaluating the model using $47$ healthy and $22$ Alzheimer's disease patients from the ADNI dataset to gauge the generalization capabilities of the proposed hybrid model. However, the authors did not perform any regional analysis to find specific brain areas related to the Alzheimer's disease. 

Finally, a recent step in the direction towards voxel-based CNN is \citet{Hong2020}, which proposes a $3$D-CNN architecture for the prediction of brain age in children from routine brain MRI scans. The authors used data augmentation strategies to extend the data size used to train the deep neural network. Moreover, to evaluate the correlation among the adjacent slices of the brain scans, the authors sliced the $3$D scans into multiple $2$D slices followed by training via a $2$D CNN model. However, the performance of the $2$D-CNN model was found to be lower when compared to the $3$D-CNN model. This signifies the fact that the correlation among the adjacent slices is of utmost relevance for the brain age prediction. Interestingly, \citet{Hong2020} unveiled that predictions for children age under $2$ years are more accurate than the ones elicited for children aged over $2$ years.
 
\subsubsection{VGGNet}

Because of the popularity of DL networks, researchers are attracted to pretrained DL networks to get better performance results for brain age estimation. VGGNet \cite{simonyan2014very} is the pretrained DL network having the fixed size convolutional layer $(3 \times 3)$, which allows modeling small details of the input image. Moreover, the number of convolutional filters gets doubled at each stack of the convolutional layer. We now discuss on slice-based and voxel-based VGGNet models used for brain age estimation:

\begin{itemize}[leftmargin=*]
\item Slice-based VGGNet models:
\end{itemize}

To the best of our knowledge, the work in \cite{huang2017age} is the only one investigating the performance of 2D VGGNet for brain age estimation. In doing so, the work relies on VGGNet for extracting 2D slices from 3D volumetric images having the highest mutual information to increase the speed of the network. A downside of this work is the use of only channel normalization for the preprocessing of MRI images, ignoring motion artefacts, linear registration and bias correction in the preprocessing that may lead to inaccurate results. Furthermore, authors did not compare the performance of VGGNet with other pretrained networks.

\begin{itemize}[leftmargin=*]
\item Voxel-based VGGNet models:
\end{itemize}

As opposed to slice-based strategies, voxel-based VGGNet models have been more frequently used in the literature. To begin with, \citet{feng2020estimating} investigated a large-scale heterogeneous MRI dataset for brain age estimation using a 5-layer 3D VGGNet model. In addition, the authors also conducted a region-wise age analysis, and demonstrated a negative association between frontal lobe measures like cortical thickness and the age difference. Further, the association of predictive age difference and Benton Face Recognition scores (BFRT) is used to measure the baseline visual memory in evaluating cognitive functions. In conclusion, the study provides quantitative and region-based analysis for predicting brain age and other neuroimaging applications. However, it only evaluated the BFRT score to correlate the age deviation with the cognitive decline, which is not specific to a particular disease. Another limitation is the inclusion of less data from very old individuals, excluding data for individuals less than 18 years old.

Another more recent contribution exploring voxel-based VGGNets is \cite{dinsdale2021learning}, which explored 3D VGGNet models using the T$1$-weighted UK Biobank dataset. They examined the correlation between the predicted age difference and $8,787$ non-image variables reporting on the lifestyle, medical and physiological parameters, mental health self-report, and medical history of the patient. The study uncovered a mild correlation between the age difference and the image-derived phenotype (IDP) of multi-modalities. Attention gates are also investigated with linearly and non-linearly registered images to focus on the age-affected region in the brain and suppress non-affected regions. Linearly registered MRI scans show the subtle cortical changes which are not visible in nonlinear registered scans, proving the efficiency of the utility of linear registered images for detecting cognitive decline. As a drawback of this study, we highlight the use of IDPs to investigate the relation between the other modalities and age difference, as IDPs consist of reduced voxel-wise information when compared to T$1$-weighted $3$D scans.

An interesting alternative was proposed by \citet{peng2021accurate}: a lightweight, simple fully-connected convolutional network (SFCN) inspired by the VGGNet model. SFCN uses dropout, voxel shifting and mirroring to improve the performance results. In exchange, the model requires more training time than other modeling alternatives at the time, which is more than $50$ hours using 2 specialized P100 GPU cards. In addition, the study did not investigate potential associations between age difference and any health condition of the subject. 

Finally, we stop at the work by \citet{jiang2020predicting}, which segmented T$1$-weighted images using the CorticalParcellation-Yeo$2011$ reference map, yielding $7$ regions (cortex networks) modeled each by a 3D VGGNet model. The results showed less mean absolute error using the frontoparietal network, dorsal attention network, and default mode network. They also investigated the association of GM to brain age-related changes using Pearson's correlation. However, the use of only $7$ cerebral cortex networks ignores age effects emerging from fine-grained subnetworks and subcortical networks.

\subsubsection{ResNet}

As explained in Subsection \ref{ssec:cnn_resnet_vgg}, ResNet overcomes training issues derived from vanishing gradients by using hierarchical stacks of residual blocks. We now review works in which brain age estimation has been approached by ResNet-based deep learning methods:

\begin{itemize}[leftmargin=*]
\item Slice-based ResNet models:
\end{itemize}

Our literature collection only discovered a single work prospecting the adoption of slice-based ResNets for brain age estimation. \citet{shi2020fetal} explored the prediction of the fetal brain age using ResNet as the basic building block of a CNN trained over T$2$-weighted images. Furthermore, an attention map was extracted from the last stage of the CNN network. As a result, the study found out an association between the predicted age difference and fetal anomalies like small head circumference and malformations. The main limitation of this work is the use of transverse MRI scans for acquiring the T$2$-weighted scans of fetal, but some of the brain anomalies like corpus callosum and other midline brain structures are poorly visible in the transverse plane. Another weakness is the use of the last menstrual period as the ground truth of gestational age, which may be inaccurate.
 
\begin{itemize}[leftmargin=*]
\item Voxel-based ResNet models:
\end{itemize}

The seminal work in \citet{jonsson2019brain} proposed the use of a $3$D ResNet based multimodal model for brain age estimation, which takes T$1$-weighted, WM, GM and Jacobian map as the input to each CNN network. Also, sex and scanner information are added at the last layer to improve the performance of the model. To avoid randomly initializing the CNN network, the ensemble model is first trained on the Icelandic dataset, then fine-tuned using transfer learning over the IXI dataset. Then the model is assessed on UK Biobank dataset, and final decision is made by using majority voting among the trained networks. A genome-wide association study is also done to link the predicted age difference to two main genome sequence: rs$1452628$-T and rs$2435204$. The study also show the negative association of the predicted age difference and neuropsychological time tests. A clear downside of the proposal made in this study is its computational complexity derived from the use of an ensemble of 3D CNN networks.

Shortly thereafter, \citet{kolbeinsson2020accelerated} resorted to a voxel-based approach to investigate the extent to which six brain regions contribute to brain ageing, including the left amygdala, right hippocampus, left cerebellar, left insular cortex and the left crus and vermis. To this end, a permutation importance method was used. The authors explored the age differences associated with hypertension, multiple sclerosis, systolic and  diastolic blood pressure measurement, and Type-I and II diabetes. Unfortunately, the use of permutation based importance quantification is limiting for regional analysis, as it shows that boundary affects if images are not correctly aligned. 

More recently, two contributions have made further use of residual network architectures. The first is \cite{fisch2021predicting}, which proposed a ResNet based $2$-layer 3D CNN architecture with minimal preprocessing of T$1$-weighted images and used transfer learning to extract features. They used only brain extraction and image cropping as the preprocessing steps to reduce the computational complexity of the proposed model. Training is performed over the GNC dataset and evaluated using BiDirect study, FOR$2107$/MACS, and the IXI datasets using transfer learning to inspect the generalization potential of the model. Since $3$D CNN was only in use for age prediction, no comparison of the results to other pretrained deep learning networks was done. Furthermore, the correlation of age difference with brain regions was not investigated. 

The other recent contribution related to voxel-based ResNet models is \cite{Ning2021}, which used this architecture for identifying genetic factors linked to the brain aging. The study, based on the genome-wide association study with predicted brain age, revealed that single nucleotide polymorphisms (SNPs) from four independent locations are significantly correlated with brain aging. However, the work only evaluated the genetic factor while ignoring other aspects like lifestyle habits or diseases. For example, heavy smoking and the consumption of alcohol are known to accelerate brain aging \cite{ning2020association}. Likewise, diseases like diabetes and schizophrenia play a catalytic role in brain aging as well \cite{franke2013advanced,schnack2016accelerated}. By contrast, physical exercise improves the general health and slows the brain aging \cite{kramer2006exercise,larson2006exercise}. 

\subsubsection{Ensemble Deep Learning}
Ensemble deep learning combine multiple networks to achieve improved robustness and predictive performance, at the cost of a generally higher training complexity. Researchers started to embrace ensemble deep learning to achieve better results for brain age estimation. A deeper analysis of the works in this direction is done in what follows:

\begin{itemize}[leftmargin=*]
\item Slice-based Ensemble Deep Learning models:
\end{itemize}

Two contributions from the reviewed literature fall within this first category. To begin with, \citet{Ballester2021} proposed slice-level brain age prediction via CNNs and linear regression. Instead of using a whole brain image, the authors used slice-level predictions to identify the brain regions that most contribute to the brain age gap. Moreover, authors evaluated how the predictions are affected by factors like the slice index and plane, age, sex and MRI data collection site. The study revealed that specific brain slices have more impact on the prediction error. A stratified partitioning of the train and test datasets was shown to remove the site effects and minimize the sex effects. Also, the overall error of the models is affected by the choice of the MRI slice. 

The second work is \cite{Hwang2021}, which used an ensemble of CNNs to estimate brain age from routine T$2$-weighted spin-echo brain MRI scans. The study suffers from the following limitations \cite{Wood2022}: 1) it averaged predictions across all slices, however, not all slices may contribute equally to the final prediction, potentially resulting in subpar performance; 2) each slice is treated independently from the rest of slices, which precludes the modelling of non-linear interactions among the axially separated features. There are several shortcomings of brain age models that are not developed for routine clinical MRIs. To begin with, models are mostly trained over high-resolution volumetric T$1$-weighted scans with isotropic or near-isotropic voxels. However, these type of scans are rarely the part of routine examinations. Secondly, most models are trained and evaluated on the datasets which follow precise imaging protocols and participant inclusion criteria. Contrarily, routine imaging scans in hospitals are heterogeneous in practice due to varying scanner vendors, different imaging protocols and heterogeneous patients from multiple sites. Thirdly, state-of-the-art brain age models rely on computationally intensive preprocessing techniques such as spatial normalization, skull stripping and bias field corrections.

\begin{itemize}[leftmargin=*]
\item Voxel-based Ensemble Deep Learning models:
\end{itemize}

Four contributions in the reviewed literature can be classified as voxel-based ensembles of deep learning models. The first to highlight is \cite{Kuo2021}, which empirically showed the relation between input feature sets and the predictive performance of brain age estimation methods under the framework of conventional machine learning models. The authors showed that the combination of multiple input feature sets and objective-specific functions via an ensemble deep learning framework results in an improved performance. Similarly, \citet{couvy2020ensemble} utilized an ensemble of $7$ classifiers, which were trained over $3$D GM and WM maps and vertex-wise measurements of surface area. They concluded that the best performance was achieved by using ensemble deep learning with tuneable weights estimated by using linear regression. Further, a combination of the ensemble predictions using random forest improves the performance of the proposed model even further. Unfortunately, as in any other ensemble deep learning method, a penalty in computational complexity yields from the need for training $7$ different networks. 

Another example of voxel-based ensemble deep learning is \cite{levakov2020deep}, which proposed an ensemble of multiple 3D CNN networks on $10,176$ subjects for chronological age prediction. The model achieved a MAE of $3.07$ years over unseen data. They located cerebrospinal fluid cavities as an atrophy marker for age prediction. Furthermore, the authors aggregated several explanation maps by averaging to form a population-based map and highlighted the ventricles and cisterns as potential biomarkers related to early aging. This work considered only cross-sectional data instead of longitudinal data; as a result, only between-subject age variations were accounted for, leaving aside possible trajectories within the same subject. \citet{Hofmann2021} proposed multilevel ensemble models towards the interpretability of the models in neuroimaging, showing better performance than their individual base learners. The authors concluded that the voxels in and around the ventricles and at the border of the brain to meningeal areas contributed strongly to the brain age. Also,  voxels covering cortical sulcal structures appeared to be more relevant in older subjects than in the younger ones. Finally, \citet{poloni2022deep} developed two brain age estimation CNN models focused on hippocampal age by using T$1$ brain MRI scans. Authors used data augmentation techniques that provided ample amount of data to train the model.This approach explores an association of prediction errors with different subjects such as cognitively normal, AD and MCI subjects. They performed statistical analysis of delta estimation error among baseline groups. In addition, authors correlated the results with clinically measured values and found a negative correlation. This research provided the inference time of $0.12$ seconds and the overall processing time mentioned is below seven minutes.

\subsubsection{Miscellaneous Deep Learning Models}

Part of the literature has glimpsed at other pretrained deep learning networks for brain age estimation, including Alexnet \cite{krizhevsky2012imagenet}, U-Net \cite{ronneberger2015u}, DenseNet  \cite{huang2017densely}, Autoencoder \cite{vincent2010stacked}, Inception-V1 \cite{szegedy2015going}, and EfficientNet. For completeness these works are briefly discussed next:

\begin{itemize}[leftmargin=*]
\item Slice-based miscellaneous Deep Learning models:
\end{itemize}

\citet{lin2021utilizing} proposed 2D AlexNet \cite{krizhevsky2012imagenet, iandola2016squeezenet} for feature extraction, principal component analysis (PCA) for feature selection and relevance vector machine with polynomial kernel for classification using GM images extracted from T$1$-Weighted images. They tested the Alzheimer's patient data and found the predicted age difference of $3.13$ years of Mild Cognitive Impairment (MCI) patients and $6.08$ years for Alzheimer's Disease (AD) patients. However, authors demonstrated the results only using Alexnet \cite{krizhevsky2012imagenet, iandola2016squeezenet} pretrained model, which is a black-box and lacks reliable explanation. Furthermore, the authors did not investigate the performance on other pretrained networks to comparatively assess the efficiency of their AlexNet-based proposal.

An interesting advance is the work by \citet{lombardi2021explainable}, which presented an explainable deep learning networks with morphological features extraction. The authors investigated two morphological descriptors -- SHAP \cite{lundberg2017unified} and LIME \cite{ribeiro2018anchors} -- for finding reliable biomarkers for aging. The experimental analysis showed the significance of SHAP as the most reliable descriptor for explaining the morphological aging process. However, to exploit 3D MRI information at the voxel level, CNN-based salience maps and layer-wise relevance propagation mapping can be used instead of SHAP and LIME  to highlight the region of interest in MRI. In addition, a correlation analysis was performed to investigate the morphological features' importance with the age. However, a quantitative analysis is still required to correlate the most important regions for age prediction.

\begin{itemize}[leftmargin=*]
\item Voxel-based miscellaneous Deep Learning models:
\end{itemize}

When using voxel-based information at their input, the diversity of miscellaneous models reported to date is higher. We begin with \citet{popescu2021u}, which used a U-Net based model for the local brain age prediction via incorporation of voxel information. The study analysed the structural differences between the MCI and AD patients. Moreover, it also evaluated the reliability of the local brain age predictions both within and between scanners. The study revealed that for local brain predicted age difference (with Cohen's $d$ values in consideration), accumbens, putamen, pallidum, hippocampus and amygdala proved to be discriminative of the subcortical regions. Moreover, the between-scanner reliability is a moderate factor. However, the drawback of this study is the requirement of a healthy population from a given site, as a site change or side scanner effects may lead to the local brain predicted age difference distribution not centred around zero. 

Another very recent work in this assorted set of contributions is \cite{Wood2022}, which developed a DenseNet-based  model for brain age estimation based on routine clinical head examinations. The use of these data ensures that the models are trained on a range of scanners and acquisition protocols, and hence, the natural representation of the routine clinical population. This study suffers from the following limitations: 1) as the model is trained on radiologically \emph{normal for age} brains, it remains unknown how the model will behave in individuals with gross abnormalities; 2) the authors attributed the interpretability of the model to brain parenchyma; however, no systematic analysis of the brain features responsible for the brain age prediction is provided.
 
\citet{lee2022deep} used a modified 3D-DenseNet \cite{huang2017densely} with Adam optimizer (with $\beta_1=0.9$ and $\beta_2=0.99$) and MAE loss function for the brain age prediction in normal aging and dementia.  The study focused on the age and modality specific maps of the CNN model which could lead to the interpretation of the brain regions contributing most to the prediction of age subgroup and modality type using occlusion sensitivity analysis. The occlusion sensitivity analysis method masks the portion of a brain and evaluates its effect on the decisions of the MAE change. The study concluded that for a younger group ($30-40$ and $40-50$ years) posterior region with a peak at posterior cingulate cortex contributed higher, while as in $50-60$ and $70-80$ years of age groups, the inferior frontal regions along with the orbitofrontal and olfactory cortex contributed predominantly. Moreover, the metabolic data proves to be more sensitive compared to MRI based data for the brain age prediction. The limitation of this study is that it only evaluated on the neurodegenerative pathology, without its evaluation on chronic systemic medical diseases and vascular diseases which could have different patterns of brain aging.

\citet{mouches2021unifying} unified the brain age prediction and age conditional template generation with a deterministic autoencoder. In age-specific templates, the authors showed  the natural age variations with an increased ventricular volume and wider sulci associated with increasing age. The same authors proposed in \cite{Mouches2022} an autoencoder-based multimodal brain age prediction model using MRI and angiography data. The authors exploited saliency maps for investigating the contribution of cortical, subcortical and arterial structures for the prediction. The study concluded that combining brain tissue and artery information improves the performance of the models. Moreover, the lateral sulcus, fourth ventricle, and medial temporal brain regions proved to be important morphological features. 
\citet{varatharajah2018a} employed a transfer learning approach via pretrained Inception-V1 \cite{szegedy2015going} based $3$D feature extractor, followed by prediction via regressor and bucketed classification. 

We end this section by examining \cite{he2022deep}, which proposed a regression model based on \emph{deep relation learning}. The basic idea behind this model is to learn the relationship between pairs of different input images. Authors considered four relations among the non-linear pairs. These relations were learned by the model simultaneously from a single neural network, and it performed two separate tasks such as relation regression and feature extraction. For the feature extraction process, EfficientNet was employed, while a transformer was applied to model the relationship between pairs of images. The performance was evaluated over different datasets with subjects aged between $20$ and $72$ years. Different performance evaluation strategies such as estimation of brain age with different images, estimation of brain age with known age references and estimation of brain age with the same pair of images were also adopted. In their experiments, $2.38$ was the lowest MAE score achieved. According to \cite{he2022deep}, the deep relation learning model can provide better generalization performance for brain age estimation with lower MAE than other state-of-the-art approaches.

\tiny
\begin{landscape}
\begin{longtable}[t]{cccC{2cm}C{2cm}C{2cm}C{2cm}L{6cm}}
\caption{Summary of the main characteristics and results of the literature reviewed in this manuscript. $\rho$ denotes Pearson’s correlation coefficient, RMSE is \emph{root mean square error}, and MAE stands form \emph{mean absolute error}.}
\label{tab:characteristics}\\
    \toprule
    & & & & & \multicolumn{2}{c}{\textbf{Performance}} & \\
\cmidrule{6-7}
    \textbf{Year} & \textbf{Ref.} & \textbf{Modality} & \textbf{Model}& \makecell{\textbf{Dataset}\\\textbf{size}} & \textbf{MAE(years)} & \textbf{Other measures} & \textbf{Database} \\
    \midrule
    2017 & \cite{Cole2017} &  T1 & 3D-CNN & 2,001 & $4.16$ & r$=0.96$, RMSE$=5.31$ & Brain-Age Normative Control (BANC) \\ 
    \midrule
    2017 & \cite{huang2017age} &  T1 & VGGNet & 1,099 & $4$ & -- & Aoba brain image research center and Sendai Tsurugaya project\\ 
    \midrule
    2018 & \cite{varatharajah2018a} & - & 3D CNN with regression and bucket classification & 12,988 & -- & RMSE$=5.54$ and $6.44$ & ADNI \\ \midrule
    2019 & \cite{wang2019gray} & T1 & 3D CNN & 5,496 & $4.45\pm3.59$& -- & Rotterdam Study \\ 
    \midrule
    2019 & \cite{jonsson2019brain} & T1 & 3D CNN (ResNet) & 1,264 (Icelandic), 15,040 (UK Biobank), 544 (IXI) & $3.39$ & -- & Icelandic, IXI, UK Biobank \\ 
    \midrule
    2019 & \cite{amoroso2019deep} & T1 & CNN-4L $(20,100,50,20)$ & $484$ & $4.7\pm 0.1$ & RMSE=$6.2\pm1.1$, $\rho=0.95\pm0.02$ & ABIDE, ADNI, BNU, ICBM, IXI \\
    \midrule
    2020 & \cite{feng2020estimating} & T1 & 3D CNN & 10,158 & $4.06$ & $r=0.970$& Cam-CAN, IXI, SALD, DLBS, OASIS-1, CoRR, SchizConnect, ADNI, AIBL, OASIS-2, PPMI, NIFD, BGSP and SLIM \\ 
    \midrule
    2020 & \cite{dinsdale2021learning} & T1 & 3D CNN & 19,687 & $2.71 \pm 2.10$ (female), $2.91 \pm 2.18$ (male) & -- & UK Biobank \\ 
    \midrule
    2020 & \cite{kolbeinsson2020accelerated} & T1 & 3D CNN (ResNet) & 21,382 & $2.87, 3.42$ & -- & UK Biobank \\ 
    \midrule
    2020 & \cite{lam2020accurate} & -- & 2D-CNN + RNN & 10,446 & $2.86$& RMSE$=3.61$ & UK Biobank \\ 
    \midrule
    2020 & \cite{pardakhti2020brain} & T1 & 3D CNN/SVM/GPR & $562$ & $5$ & -- & IXI \\ 
    \midrule
    2020 & \cite{shi2020fetal} & T2 & Attention-based ResNet & $659$ & $0.767$ weeks & -- & Custom data (no benchmark) \\ 
    \midrule
    2020 & \cite{jiang2020predicting} & T1 & 3D CNN (VGG13) & 1,454 & $5.55$ & -- & ABIDE, BNU, ICBM, IXI, OASIS \\ 
    \midrule
    2020 & \cite{levakov2020deep} & T1 & Ensemble of 3D CNN & 10,176 & $3.07$ & $r=0.98$ & ADNI, PPMI, ICBM,AIBL, SLIM, OASIS, CANDI, IDA, COBRE, CNP, CORR, FCP \\ 
    \midrule
    2020 & \cite{couvy2020ensemble} & T1 & Ensemble of CNN, SVM, linear unbiased predictor & 2,640 & 3.33 & -- & PAC2019 dataset \\ 
    \midrule
    2020 & \cite{Hong2020} & T1 & 3D CNN & $220$ & $67.6$ days & -- & Children’s Hospital of Nanjing Medical University \\ 
    \midrule
    2021 & \cite{bellantuono2021predicting} & T1 & 3D CNN & 1,016 & $2.19\pm0.03$ & RMSE$=2.91\pm0.03, \rho=0.890\pm0.003$ & ABIDE \\ 
    \midrule
    2021 & \cite{peng2021accurate} & T1 & Simple Fully Convolutional Network (SFCN) & 14,503 & $2.14$ & -- & UK Biobank \\ 
    \midrule
    2021 & \cite{fisch2021predicting} & T1 & 3D CNN (ResNet) & 10,691 (training) + 2,173 (validation) & $2.84 \pm 0.5$ & -- & German National Cohort \\ 
    \midrule
    2021 & \cite{lin2021utilizing} & T1 & AlexNet & $594$ & $4.51$ & $r= 0.979$ & ADNI GO, ADNI 2, IXI, OASIS \\ 
    \midrule
    2021 & \cite{lombardi2021explainable} & T1 & FCDNN & 2,638 & $2.7$ & $r=0.86$ & PAC2019 \\ 
    \midrule
    2021 & \cite{Lombardi2021} & T1 & FFDNN & 2,638 & $4.6$ & -- & PAC2019 \\ 
    \midrule
    2021 & \cite{dular2021improving} & T1 & CNN+Transfer Learning & 2,543 & $3.3$ & $r = 0.91$ & Multi-site dataset \\ 
    \midrule
    2021 & \cite{Lee} & T1 & 3D modified DenseNet & 4,127 & $3.43 \pm 0.0545$ (FDG PET), $4.2055 \pm 0.2241$ (MRI) & -- & MAYO dataset \\ \midrule
    2021 & \cite{popescu2021u} & T1 & U-Net inspired FCNN & ~3,500 & $9.751$ & -- & BANC, Dallas Lifespan Brain Study, Cambridge Centre for Ageing and Neuroscience, Southwest University Adult Lifespan Dataset,  OASIS3, AIBL, Wayne State, Local data from Imperial College London \\ \midrule
    2021 & \cite{Ning2021} & T1 & ResNet & 16,998 & $2.7$ & -- & UK Biobank \\
    \midrule 
    2021 & \cite{Hwang2021} & T2 & 3D CNN & 1,530 & $4.22$ (local), $9.96$ (IXI) & $r=0.862$ (local), $0.861$ (IXI) & Seoul National University Hospital data, IXI \\ 
    \midrule 
    2021 & \cite{Ballester2021} & -- & Modified ResNet18 & -- & $4.62$ & -- & PHOTON-AI \\ 
    \midrule
    2021 & \cite{Mouches2021} & T1 & CNN+Autoencoder & 2,118 & $4.95$ & -- & Private data (Pomerania, Germany), IXI \\ 
    \midrule
    2021 & \cite{Kuo2021} & T1 & Ensemble Deep Learning & 2,118 & $3.33$ & -- & PAC2019 \\ 
    \midrule
    2021 & \cite{Hofmann2021} & T1 & CNN & $2637$ & $3.38-5.07$ & -- & Private data \\ 
    \midrule
    2022 & \cite{Wood2022}& T2 & DenseNet121 & 23,302 & $2.97$ & -- & Data from King's College Hospital NHS Foundation Trust (KCH), Guy's and St Thomas NHS Foundation Trust (GSTT) \\ 
    \midrule
    2022 & \cite{Mouches2022} & T1 & SFCN & 2,074 & $3.85\pm2.9$ & -- & T1-weighted MRI (CNNT1) and TOF MRA (CNNTOF) datasets \\ 
    \midrule
    2022& \cite{he2022deep} & T1 & SFCN & 6,049 & $2.38$ & $\rho =0.988$ & TMGHBCH, NIH-PD, ABIDE-I, BGSP, BeijingEN, IXI, DLBS, OASIS \\
    \midrule
    2022 & \cite{poloni2022deep} & T1 & EfficientNet & 1,554 & $3.31$ & RMSE$= 4.65$& NAC, IXI and ADNI \\ 
    \bottomrule
\end{longtable}
\end{landscape}
\normalsize

\section{Challenges and future directions} \label{sec:challenges}

The literature review performed in this survey has exposed the growing interest arisen by Deep Learning models for brain age estimation over the years. Despite its relative maturity, our critical analysis of contributions to date have unveiled several research niches that still deserve further attention by the research community. We herein list such challenges, together with potential research avenues that can be traversed towards addressing them effectively in the future (we refer to Figure \ref{fig:fig_Future} for a graphical summary):
\begin{figure*} [h!]
    \centering    \includegraphics[width=\textwidth]{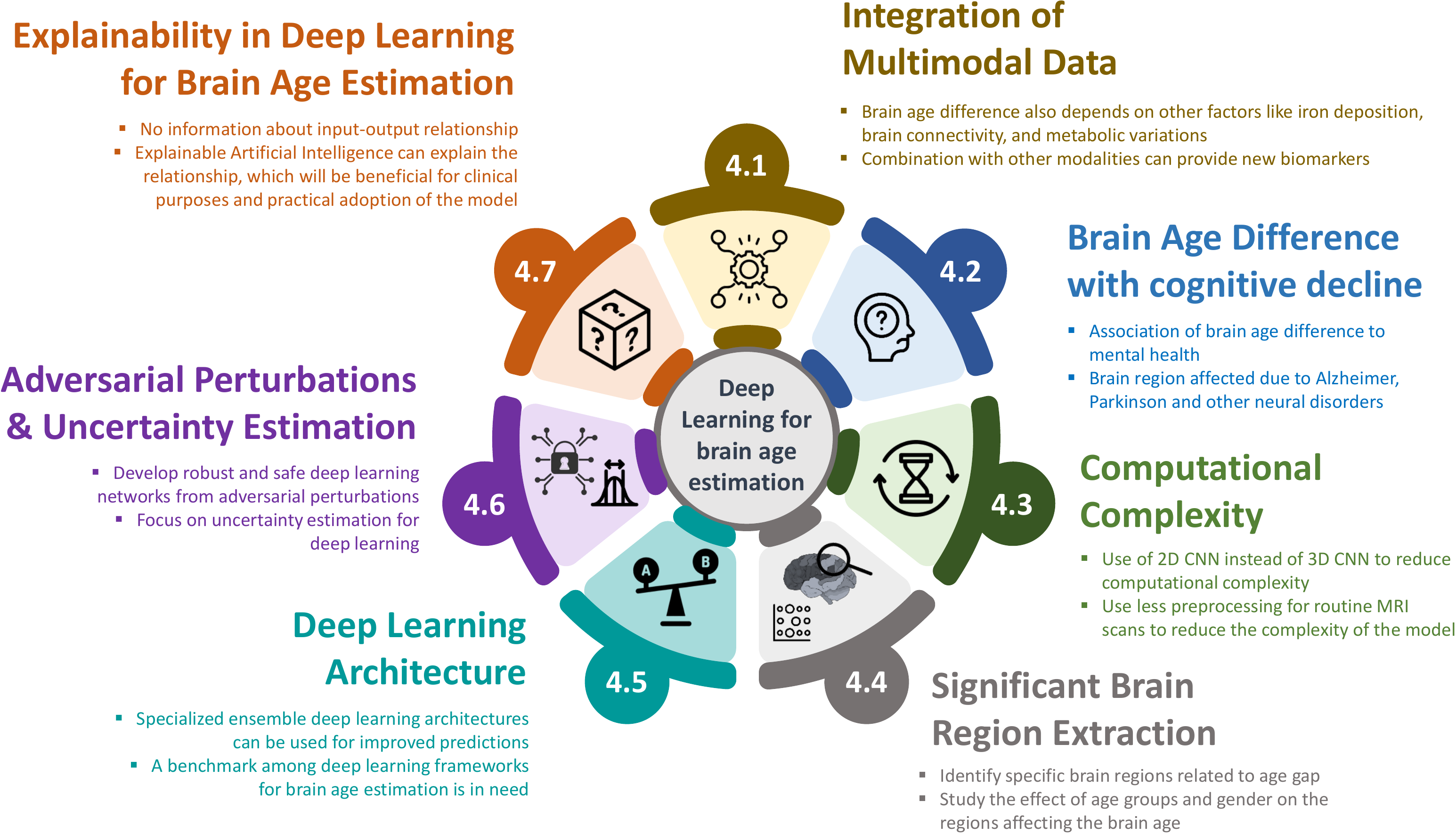}
  \caption{Graphical representation of challenges and future directions in Deep Learning for brain age estimation. Numbers refer to the subsection in Section \ref{sec:challenges} where each challenge is elaborated.}
  \label{fig:fig_Future}
\end{figure*}

\subsection{Integration of Multimodal Data }

Most brain age estimation approaches published to date used exclusively T$1$-weighted images, which can only expose structural atrophies of the brain organ. However, cognitive decline is known to stringently depend on several other factors, such as the iron deposition, brain connectivity, and metabolic variations. By virtue of their modular and highly flexible design, deep learning models are naturally prone to the integration of multiple modalities of information defined over heterogeneous domains, including space and time. Therefore, the fusion of other modalities towards a more accurate estimation of the brain will surely provide new insights and chances to explore other biomarkers that affect brain aging.

\subsection{Association of Brain Age Difference with Cognitive Decline}

In general, brain age estimation frameworks proceed by building a prediction model by modeling brain scans of cognitively healthy participants. Recent studies have documented that brain age is associated to mental health, life satisfaction and metabolic factors (e.g., diabetes) in cognitively healthy elderly \cite{sone2022neuroimaging}. On the other hand, these factors could cause a bias in estimated brain age, which should be taken into account when building a brain age estimation model from healthy data. The majority of deep learning-based brain age studies have tested their techniques on AD patients who have extensive brain atrophies. Albeit the deep learning models have shown outstanding performance on the training set and cognitively healthy subjects, their dependability should be verified on other neurological disorders with small volumetric alterations such as Parkinson’s disease, Huntington’s Disease, Epilepsy and Down Syndrome, to mention an exemplifying few.

\subsection{Computational Complexity}

A small subset of the reviewed contributions has focused on 2D CNN models, mainly aimed to reduce the computational complexity of the designed network when compared to 3D CNN modelling counterparts. 2D slices can be extracted from volumetric MRI scans based on mutual information and entropy of the slices. Furthermore, a performance comparison between the axial, coronal, and sagittal planes of MRI scans can be explored, providing a good visibility of the regions of interest. 

On the other hand, most brain age estimators rely on computationally intensive preprocessing techniques. However, only a few models have been developed for the estimation of brain age from routine MRI scans. Hence, the community should focus on developing efficient and accurate models for brain age estimation over routine MRI scans, potentially encompassing artifacts that unless properly tackled, hinder the predictability of the brain age from such data sources.

\subsection{Significant Brain Region Extraction}

The learning task underneath most brain age estimation models seeks the accurate prediction of the chronological age of the brain. These models are generally used to predict the age gap in non-healthy subjects for clinical purposes. The wider the age gap, the more abnormalities exist in the brain. However, despite the accurate prediction eventually elicited by the model, only a few studies analyze which brain regions contribute most to the estimation of the brain age. The scarce reports where such an analysis was provided have revealed that different regions of the brain are affected depending on the age group to which the subject belong. Hence, future research can be devoted towards case studies reflecting the effect of age groups and gender on brain age estimation. 

\subsection{New Deep Learning Architectures}

As evidenced in our literature review, it has not been until recently when ensemble deep learning architectures have been used to improve the brain age prediction accuracy. The most straightforward approach to induce diversity among base learners working with a given data modality  is bagging. However, there are other approaches like negative correlation learning, which can improve the performance of ensemble architectures, that remain unexplored for this particular application.

Further along this line, the community lacks a comprehensive study that benchmarks the performance of deep learning architectures on publicly available datasets, using unified metrics, evaluation protocols and several datasets. Hence, the area is in demand for a benchmark between the different deep learning approaches published for brain age estimation. This will not only contribute, in an informed fashion, to a consensus around the state-of-the-art deep learning model for accurate brain age prediction, but also stimulate further efforts towards improving such a model from different perspectives, including precision, trustworthiness and training/inference efficiency.

\subsection{Adversarial Perturbations and Uncertainty Estimation}

Deep learning has been recently found to be vulnerable to certain imperceptible perturbations called adversarial perturbations \cite{yuan2019adversarial,lyu2015unified}. Those examples corrupted with adversarial perturbations are called adversarial examples. Attacks comprising such examples unleash serious robustness challenge for deep learning methods, and have recently attracted great interest in the community. Since brain age estimation methods mainly resort to existing deep learning methods, they seldom consider their robustness against adversarial attacks. Many proposals have been proposed to enhance and/or deliver a robust and safe deep learning model \cite{zhang2021tiny,qian2022survey}. We advocate for developments tackling such robustness issues when applying deep learning models to brain age estimation.

Another important aspect of deep learning models relates to the propagation of the inherent data (aleatoric) and modeling (epistemic) uncertainty to their predicted output. Failing to provide understandable information about the confidence of a model in its produced estimation jeopardizes the adoption of the model in practical applications where decisions made from the estimations may have catastrophic consequences, as in clinical practice. Uncertainty estimation techniques aim to improve the trustworthiness of the users in the output of data-based models by informing about the confidence associated to each model \cite{kim2016examples}. For regression tasks, this information takes the form of confidence intervals, which are usually estimated for a given confidence level. Hence, prospective studies can focus on  uncertainty estimation techniques and metrics that are either model-agnostic (e.g. conformal prediction or quantile regression) or specific for deep learning models (correspondingly, evidential losses, Monte Carlo dropout or Bayesian networks), so as to increase the trustworthiness of clinical practitioners on the brain age estimated by this family of models.

\subsection{Explainability in Deep Learning for Brain Age Estimation}

In general, deep learning models are considered ``black boxes" due to the lack of an inference process that maps which components of a given input are relevant for the output value issued by the model. When modeling complex data, the performance of algorithmically transparent models (e.g. decision trees) usually lags behind that achievable by highly parametric deep neural networks. The need for techniques capable of explaining what already trained models observe in their input to predict their outputs gave birth to the landscape of post-hoc explainability methods \cite{BARREDOARRIETA202082}. Nowadays a plethora of approaches to furnish \emph{a posteriori} explanations for deep neural networks can be found in the literature, ranging from concept relevance methods to model simplification approaches and counterfactual generation. Most of these techniques have been integrated in mainstream software libraries used for the purpose. We firmly make the case for the inclusion of post-hoc explanations in future studies proposing new neural architectures and/or datasets for brain age estimation: this is necessary not only to inform decisions that may put at risk the life of patients \cite{holzinger2022information}, but also to provide medical experts with information about new biomarkers that highly correlate with brain ageing. Although exemplary attempts in this regard have been reported recently, such as the use of SHAP and LIME \cite{lombardi2021explainable}, the community should look beyond purely performance-driven research by ensuring that the model not only performs well, but also performs reliably and that can be trusted.

\section{Concluding Remarks and Outlook} \label{sec:conclusions}

This survey has reviewed brain age estimation frameworks encompassing deep learning architectures at their core. Essentially, the estimation of the brain age from neuroimaging data can be regarded as a predictive modeling task. The high dimensionality of neuroimaging MRI scans and other modalities called for the adoption of different deep learning models for the early diagnostic of several disorders directly or indirectly related to brain age estimation. This review has attempted at collecting them in their entirety, examining in depth how the brain age estimation problem has been approached in each published study, which modeling patterns have been followed to solve them effectively, reported levels of predictive performance and relevant architectural choices. As a result, a taxonomy has been proposed to arrange and organize all the literature in a systematic manner, serving as a guide for the critical assessment of the strong points and weaknesses of the reviewed works. As a consequence of our in-depth review of the literature, we have also enumerated several research directions in brain age estimation to drive future efforts in this topic. We hope this manuscript enthrones as an useful reference for researchers and practitioners working on brain age estimation via deep learning architectures, as well as for newcomers willing to join forces around this exciting research area. 


\section*{Acknowledgment}
This work is supported by National Supercomputing Mission under DST and Miety,  Govt. of India under Grant No. DST/NSM/R\&D\_HPC\_Appl/2021/03.29. We are grateful to the Indian Institute of Technology Indore for  the support and facility  offered for the research. J. Del Ser would like to thank the Spanish \emph{Centro para el Desarrollo Tecnologico Industrial} (CDTI, Ministry of Science and Innovation) through the ``Red Cervera'' Programme (AI4ES project), as well as by the Basque Government through the ELKARTEK program and the Consolidated Research Group MATHMODE (ref. IT1456-22).

\bibliographystyle{IEEEtranN}
\bibliography{refs,references}

\end{document}